\documentclass[11pt,twoside]{article}
\usepackage{asp2010}

\resetcounters

\begin{document}

\title{Absorption and emission line studies of gas in the Milky Way halo}
\author{N. Ben Bekhti$^1$, P. Richter$^2$, B. Winkel$^1$, J. Kerp$^1$, P. Kalberla$^1$, U. Klein$^1$, and M. T. Murphy$^3$}
\affil{$^1$Argelander-Institut f\"{u}r Astronomie (AIfA), University of Bonn,\\  Auf dem H\"{u}gel 71, D-53121 Bonn, Germany}
\affil{$^2$Institut f\"{u}r Physik und Astronomie, University of Potsdam, Haus 28, Karl-Liebknecht-Str. 24/25, D-14476 Potsdam, Germany
}
\affil{$^3$ Centre for Astrophysics \& Supercomputing, Swinburne University of Technology, Hawthorn, Victoria 3122, Australia} 

\begin{abstract}
We perform a systematic study of physical properties and distribution of neutral and ionised gas in the halo of the Milky Way (MW). Beside the large neutral intermediate- and high-velocity cloud (IVC, HVC) complexes there exists a population of partly ionised gaseous structures with low-column densities that have a substantial area filling factor. The origin and nature of these structures are still under debate. We analyse the physical parameters of the MW halo gas and the relation to quasar (QSO) metal-absorption line systems at low and high redshifts. For this purpose we combine new \ion{H}{i} 21-cm data from the EBHIS and GASS surveys with optical quasar absorption line data to study the filling factor and distribution of these gaseous clouds in the halo at \ion{H}{i} densities below 10$^{19}$\,cm$^{-2}$. This study is important to understand the evolution of the MW in particular and the gas accretion mechanisms of galaxies in general. 
\end{abstract}

\section{Introduction}
High-resolution emission and absorption line measurements have demonstrated that galaxies at low and high redshift are surrounded by large amounts of neutral/ionised gas with a broad range of physical parameters \citep[e.g.,][ and references therein]{fraternalietal_07, sancisietal2008}. The properties of the extraplanar gas are predominantly determined by the accretion of metal-poor gas from intergalactic space and the outflow of metal-enriched gas from star forming activities \citep[e.g.,][]{sembach_wakker_savage_richter_etal03, fraternaliandbinney06}.

The most prominent Galactic halo structures are the IVCs and HVCs which represent clouds of neutral atomic hydrogen seen in 21-cm emission at radial velocities inconsistent with a simple model of Galactic disk rotation. Beside the massive IVC/HVC complexes there also exists a population of low-column density absorbers with \ion{H}{i} column densities of 
$N_\mathrm{HI} < 10^{19}$\,cm$^{-2}$. 

One key aspect for our understanding of the nature and origin of these halo clouds and their role for the evolution of the MW is the determination of metal abundances of these clouds. 
An efficient way to study parameters and distribution of these systems over a large range of column densities (at low and high redshifts) is to combine absorption line measurements in the direction of quasars and \ion{H}{i} 21-cm radio observations \citep{richter_sembach_wakker_savage_tripp_kalberlaetal01, richterwestmeierbruens05, benbekhtietal08}.

While there are a large number of recent absorption studies on the nature of IVCs and HVCs and their role for the evolution of the MW, relatively little effort has been made to investigate the connection between the Galactic population of these clouds and the distribution and nature of intervening metal-absorption systems from halos of other galaxies seen in QSO spectra. Especially, \ion{Mg}{ii} and \ion{C}{iv} absorption line systems \citep[e.g.,][]{charltonetal00, boucheetal06} and their relation to galactic structures suggest rather complex absorption characteristics of these structures indicating the multi-phase nature of gas in the inner and outer halo of galaxies. 

In our project we systematically analyse low-column density gaseous structures in the inner and outer halo of the MW detected in \ion{Ca}{ii} and \ion{Na}{i} absorption towards QSOs along 402 sight lines. Our study allows to directly compare the observed column density distribution of gas in the MW halo with the overall column density distribution of intervening metal absorbers toward QSOs. 

Studying the complex interplay of gas exchange between galactic disks and halos (infall, outflow, merging) and the analysis of the properties are crucial to understand the role of this gas for the evolution of galaxies.

\section{Data}\label{secdata}

Our analysis is based on 402 archival optical spectra of background QSOs obtained with the Ultraviolet and Visual Echelle Spectrograph (UVES) at the ESO/Very Large Telescope. The spectra used to search for the \ion{Ca}{ii} and \ion{Na}{i} lines have a spectral resolution of $R \approx 40000-60000$, corresponding to approximately $6.6\,\mathrm{km\,s}^{-1}$\,FWHM. 

The absorption measurements were complemented with single-dish \ion{H}{i} 21-cm data obtained from the Effelsberg-Bonn HI Survey \citep[EBHIS,][]{winkeletal2010} and the Galactic All-Sky Survey \citep[GASS,][]{mcClureetal2009, kalberlaetal_2010}. 
The sizes of the telescope beams are $\sim9\arcmin$ (EBHIS) and $\sim14\arcmin$ (GASS). The velocity resolution is about $1.3\,\mathrm{kms}^{-1}$ and $1\,\mathrm{kms}^{-1}$, respectively. The rms is about 90\,mK for EBHIS and 57\,mK for GASS.

Up to now five sight lines were also observed with the Westerbork synthesis radio telescope (WSRT) and the Very Large Array (VLA) to search for small-scale structures embedded in the observed diffuse gas.

\begin{figure} 
\center
\includegraphics[height=0.95\textwidth, angle=-90]{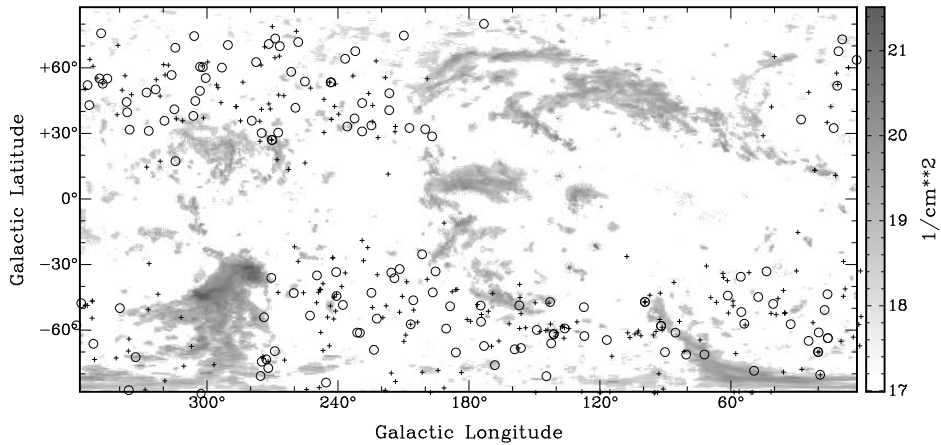}\hfill
\caption{HVC-all-sky map by Westmeier (2007) based on the LAB survey \citep{kalberla05}. The symbols mark the position of the 402 sight lines observed with UVES, GASS and EBHIS. Circles show sight lines with and crosses positions without a detection.}
\label{fig_allsources} 
\end{figure} 

\section{Observational Results}\label{secobsresults}

Figure\,\ref{fig_allsources} shows an all-sky HVC map based on the LAB-survey \citep{kalberla05}. We find \ion{Ca}{ii} (\ion{Na}{ii}) absorption in 122~(71)~out of 402~sight lines measured with UVES. Corresponding \ion{H}{i} emission was detected with EBHIS (GASS) in 36~(60)~cases out of 62~(103)~measured lines of sight. These results suggests that the MW halo contains a large number of such low-column density structures. 

\begin{figure}[!t]
\centering
 \includegraphics[width=0.41\textwidth,bb=19 562 310 840,clip]{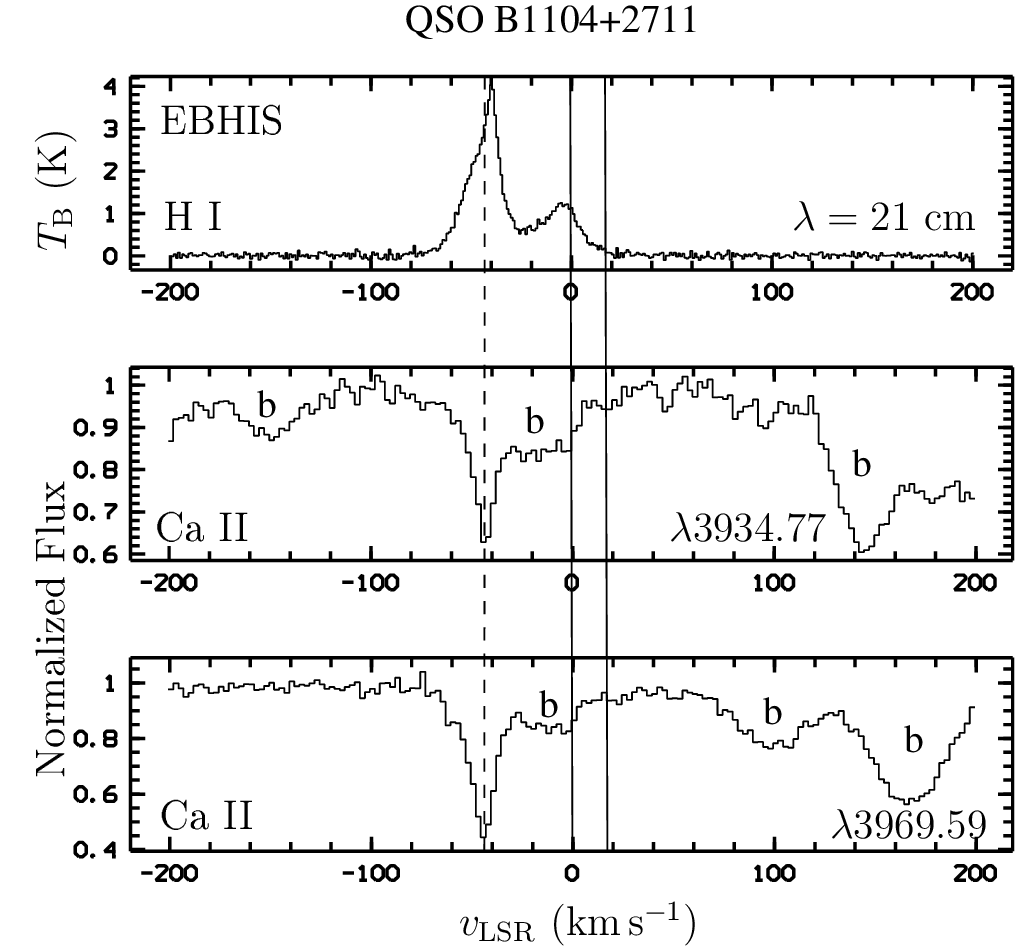}\qquad
 \includegraphics[width=0.45\textwidth,bb=22 138 613 646,clip]{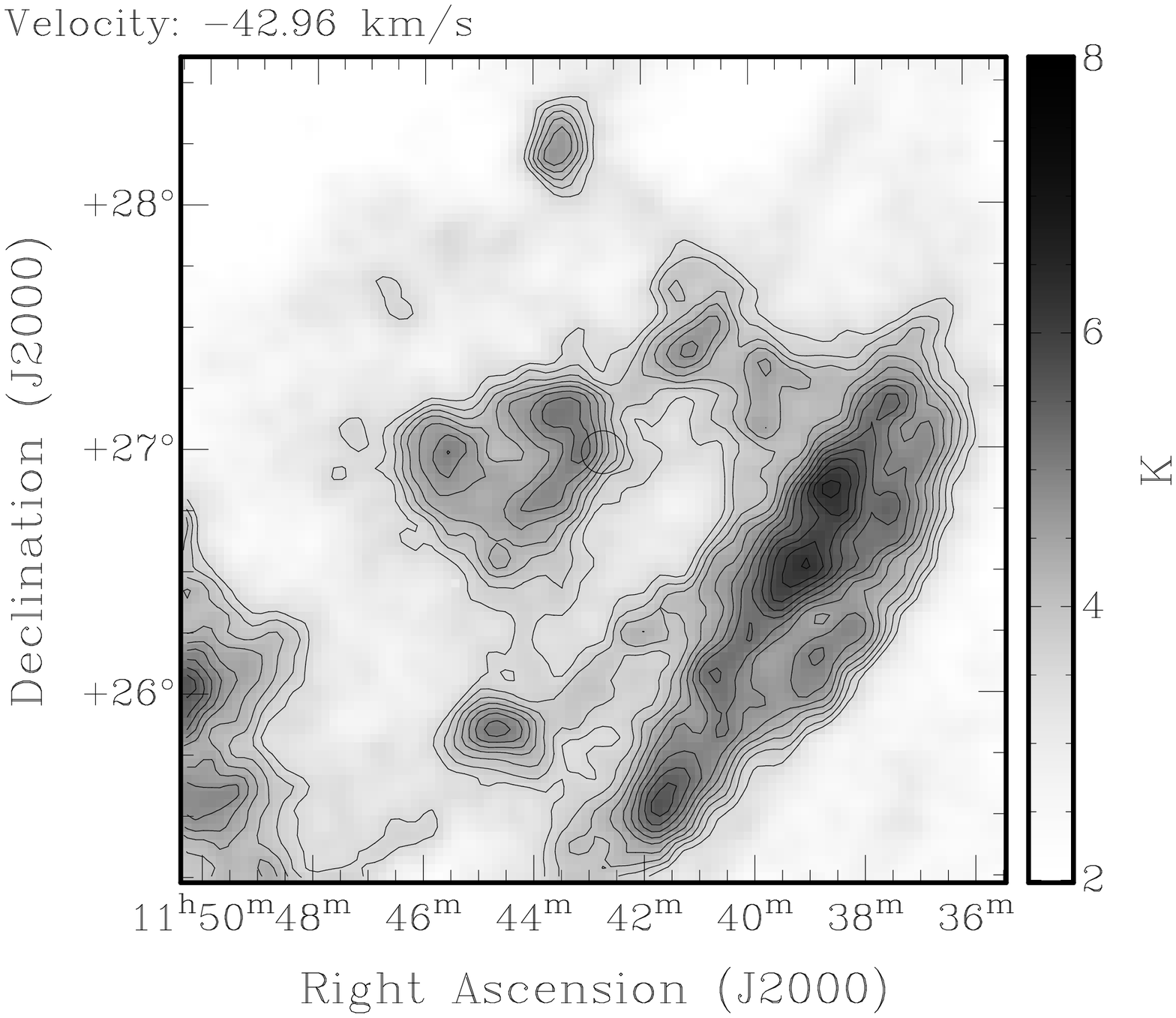}\\[-2ex]
\caption{\textbf{Left panel:} \ion{Ca}{ii} absorption and corresponding \ion{H}{i} emission spectra of QSO B1140$+$2711. \textbf{Right panel:} Channel map showing the \ion{H}{i} emission centered towards the quasar observed as part of the EBHIS \citep{winkeletal2010}. The contours start at 3.5\,K brightness temperature in steps of 0.25\,K. The circle in the center marks the size of the EBHIS beam and is positioned on the QSO sight line.} 
\label{figquasar1}
\end{figure}

For example, the line profiles of \ion{Ca}{ii} and \ion{H}{i} for one observed system are shown in Fig.\,\ref{figquasar1}. The solid lines mark the minimal and maximal LSR velocity which is expected for the Galactic disc gas according to a MW model developed by \citet{kalberlaetal2007}. 
The dashed lines indicate the location of the absorption and corresponding emission lines of the absorbers.

In many cases we observe distinct absorption lines but no corresponding IVC or HVC 21-cm emission is seen with EBHIS or GASS. This suggests that either the \ion{H}{i} column densities are below the detection limit or that the diameters of the absorbers are so small that beam smearing effects make them undetectable. 

\begin{figure}[!t]
\centering
\includegraphics[width=0.7\textwidth, bb=11 529 568 844]{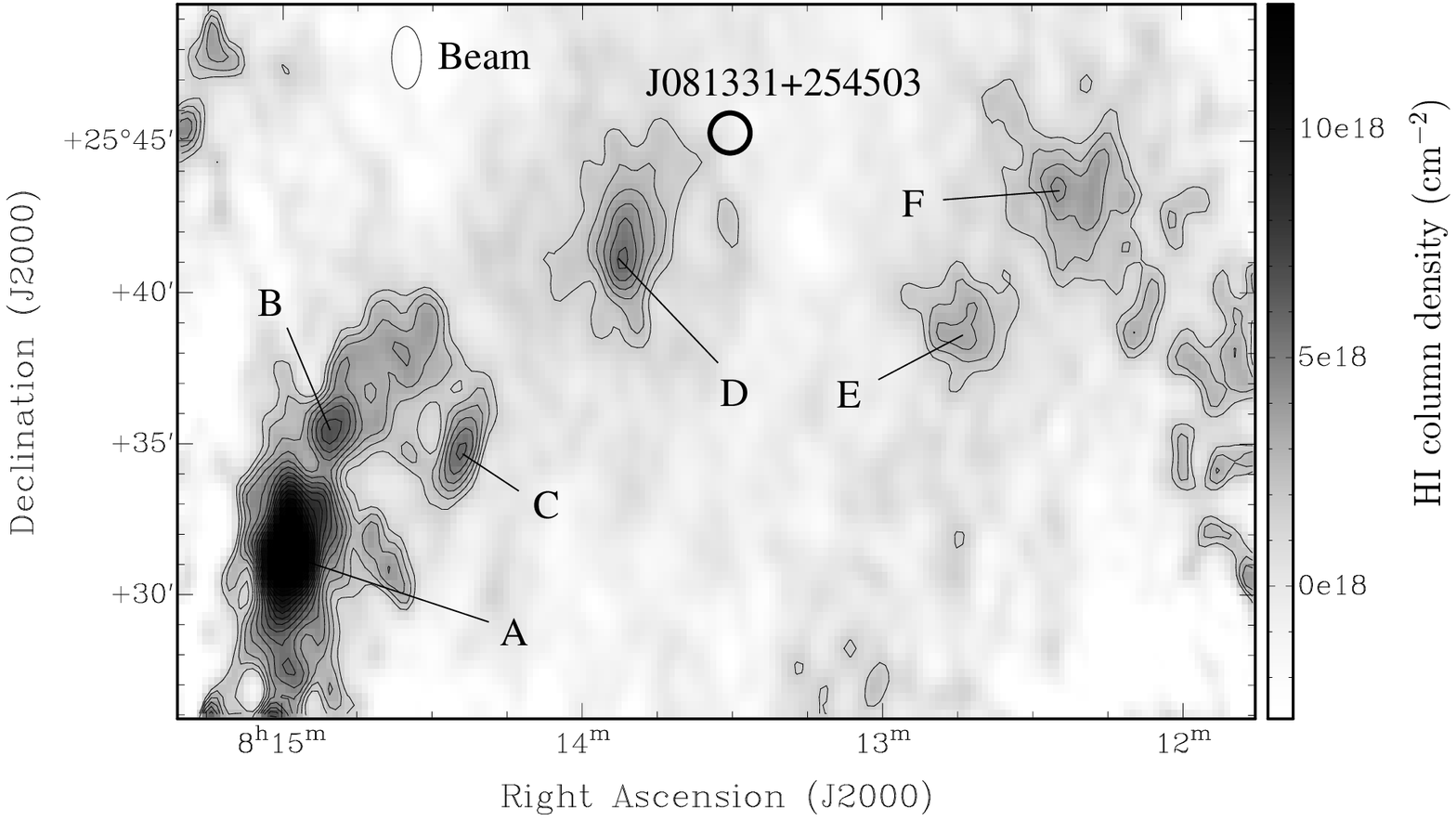}
\caption{WSRT 21-cm column density map of the IVC gas in the direction of the QSO\,J081331$+$254503 (integrated over $v_\mathrm{lsr}=-25\ldots-18\,\mathrm{km\,s}^{-1}$). The beam size of $80\arcsec \times 120\arcsec$ is shown in the upper left of the figure. The contours start at $2\sigma_\mathrm{rms}$ in steps of $\sigma_\mathrm{rms}=8 \cdot 10^{17}$\,cm$^{-2}$.}
\label{fig_J081331}
\end{figure}

The profile fitting of the optical absorption and 21-cm emission lines allows the determination of physical parameters like column densities and Doppler-parameters, $b$.
Some of our absorbers have multiple components in their spectra implying the presence of substructure. 
\citet{richterwestmeierbruens05} observed one sight line showing prominent high-velocity \ion{Ca}{ii} and \ion{Na}{i} features with UVES, Effelsberg and the VLA. The VLA resolves the HVC into several compact, cold clumps. To further investigate whether these small-scale structures are a common phenomenon we obtained additional high-resolution \ion{H}{i} data using VLA and WSRT \citep{benbekhtietal2009}. In all four observed directions we found small-scale structure embedded in the more diffuse component of the absorbing cloud \citep[for example see Fig.\,\ref{fig_J081331} and ][]{benbekhtietal2009}. The clumps have column densities of $N_\mathrm{HI}= 10^{18} \ldots 10^{19}$\,cm$^{-2}$ and angular diameters of $\Phi \leq 5\arcmin$. The measured line widths of $\Delta v_\mathrm{FWHM}= 2 \ldots 3$\,km\,s$^{-1}$ are very small resulting in upper kinetic temperature limits of $70 \leq T_\mathrm{max} \leq 3700$\,K. Furthermore, it is remarkable that the smallest spatial structures observed along the four sightlines are of similar size as the synthesised WSRT and VLA beams. Therefore, it is possible that the clumps are still not resolved and contain structures on even smaller scales.

\section{Statistical results}\label{secstatresults}

\begin{figure}
\centering
 \includegraphics[width=0.5\textwidth]{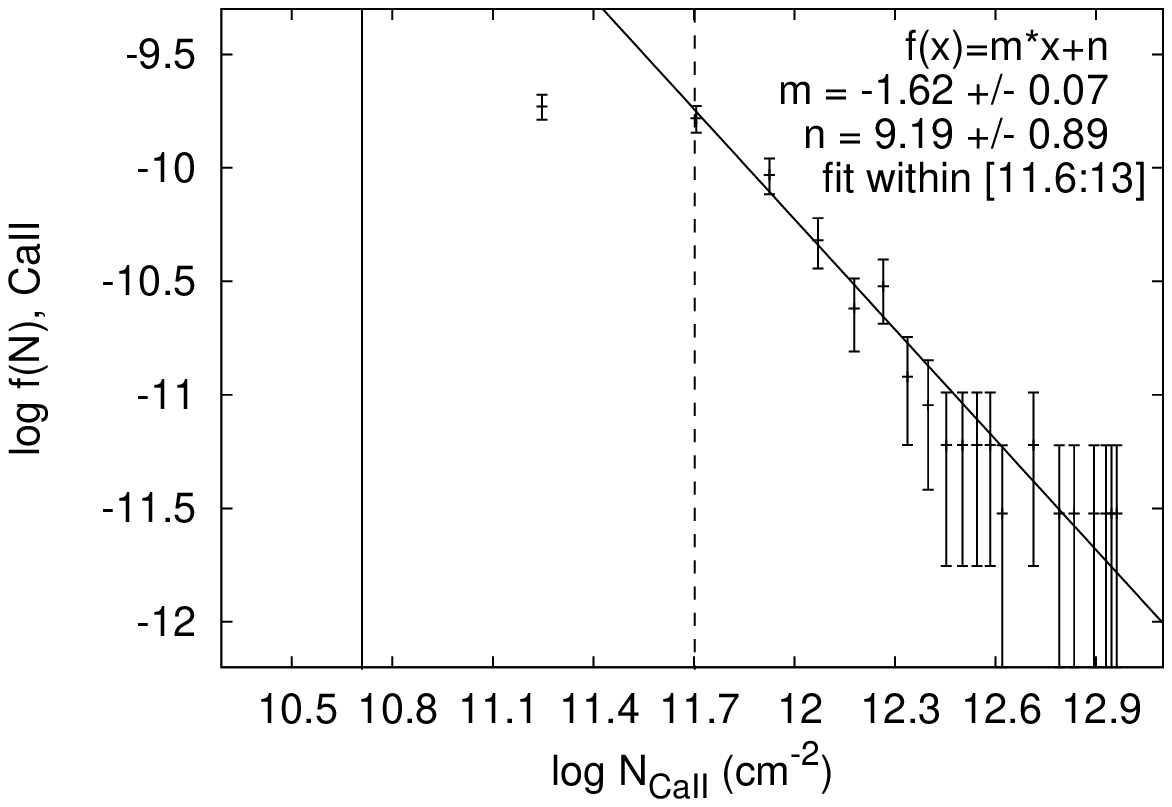}\hfill
 \includegraphics[width=0.5\textwidth]{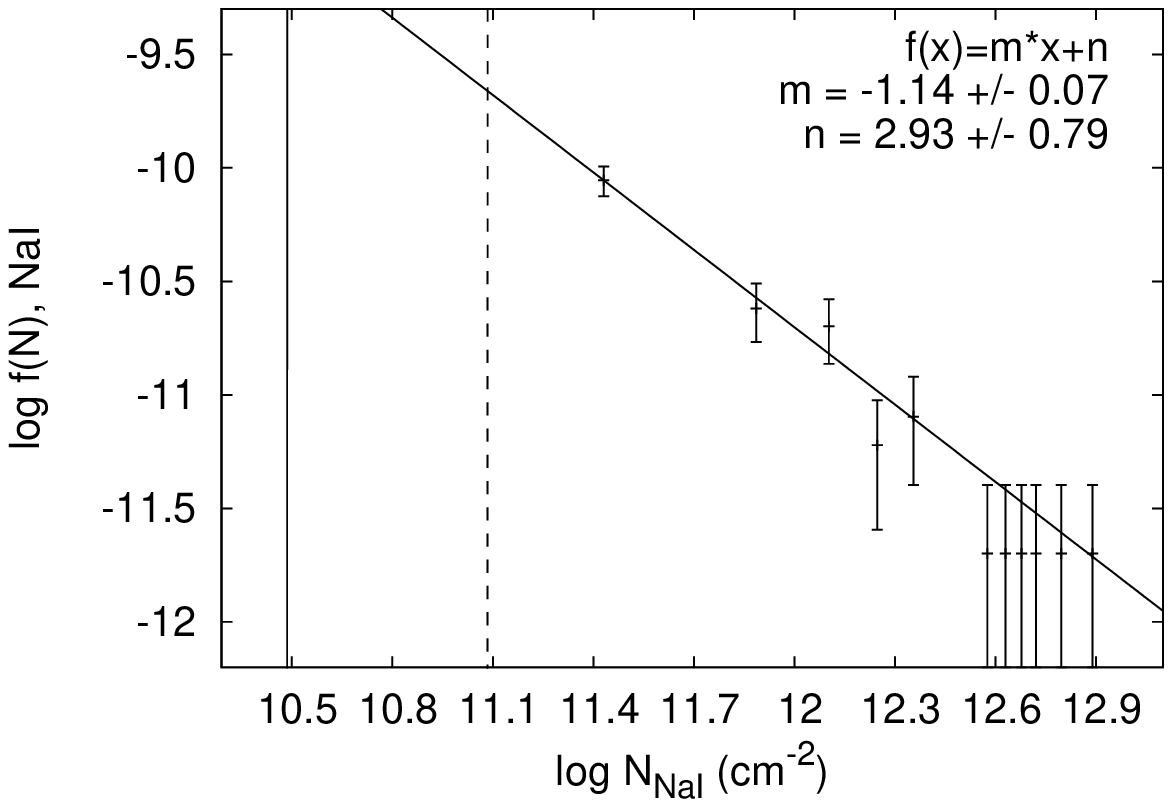}\\[-3ex]
\caption{\ion{Ca}{ii} and \ion{Na}{i} column density distribution functions.} 
\label{figcdd}
\end{figure}

Following \citet{churchillvogtcharlton03} we define the column density distribution (CDD) function $f(N)=\frac{m}{\Delta N}$ where $m$ is the number of absorbers in the column density interval [$N+ \Delta N$]. Fig.\,\ref{figcdd} shows the results for \ion{Ca}{ii} and \ion{Na}{i}. The \ion{Ca}{ii} and \ion{Na}{i} column densities for $\log N_\mathrm{CaII} > 11.6$ and $\log N_\mathrm{NaI} >11 $ follow a power law $f(N)=C N^{\beta}$ with the slopes $\beta=-1.6 \pm 0.1$ (\ion{Ca}{ii}) and $\beta=-1.1 \pm 0.1$ (\ion{Na}{i}). The vertical solid lines indicate the median detection limit of our sample and the dotted lines represent the spectrum with the highest noise level belonging to the worst detection limit. The flattening of the \ion{Ca}{ii} distribution towards lower column densities is most likely caused by a selection effect.

Churchill et al. derived the CDD function for strong \ion{Mg}{ii} systems in the vicinity of galaxies at redshifts $z=0.4 \ldots 1.2$ and found a slope of $\beta=-1.6 \pm 0.1$ which is similar to the slope of our \ion{Ca}{ii} CDD. This suggests that \ion{Ca}{ii} and \ion{Mg}{ii} probe similar gaseous structures in the environment of galaxies.

\begin{figure}
\centering
 \includegraphics[width=0.45\textwidth]{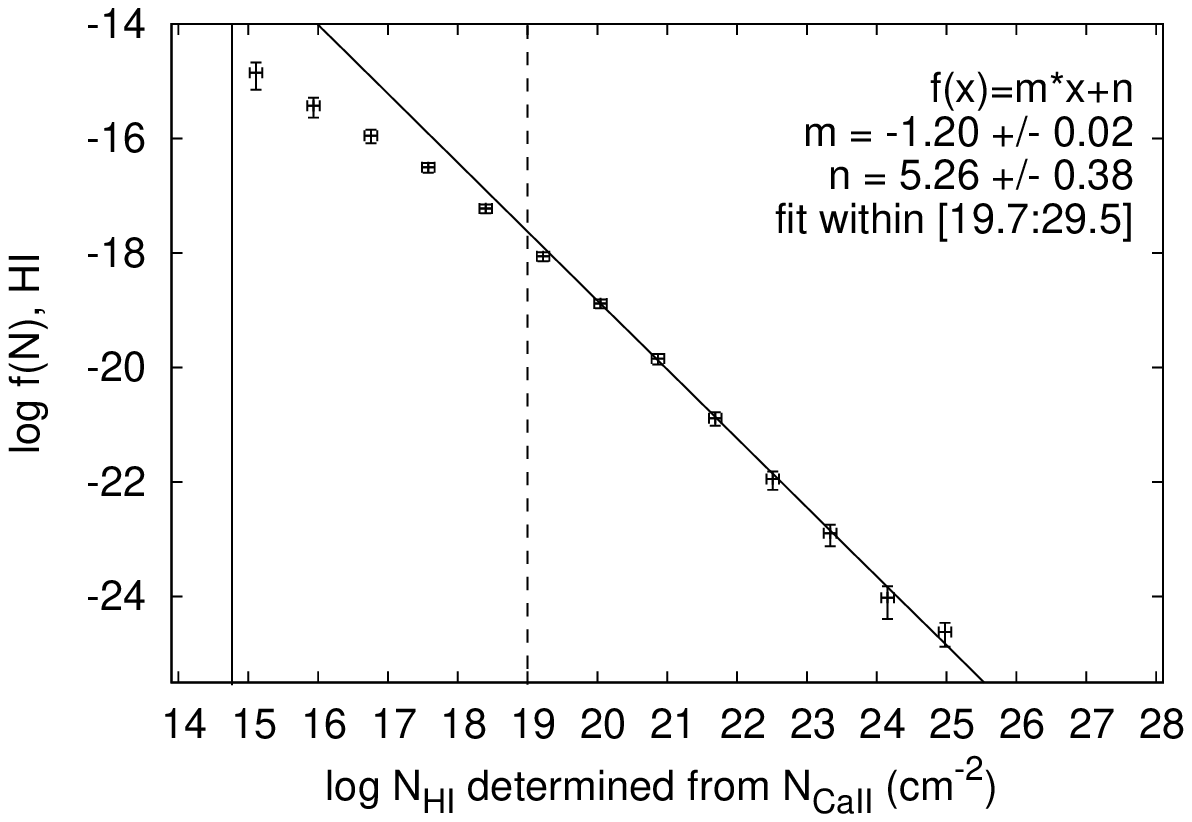}\qquad
 \includegraphics[width=0.45\textwidth]{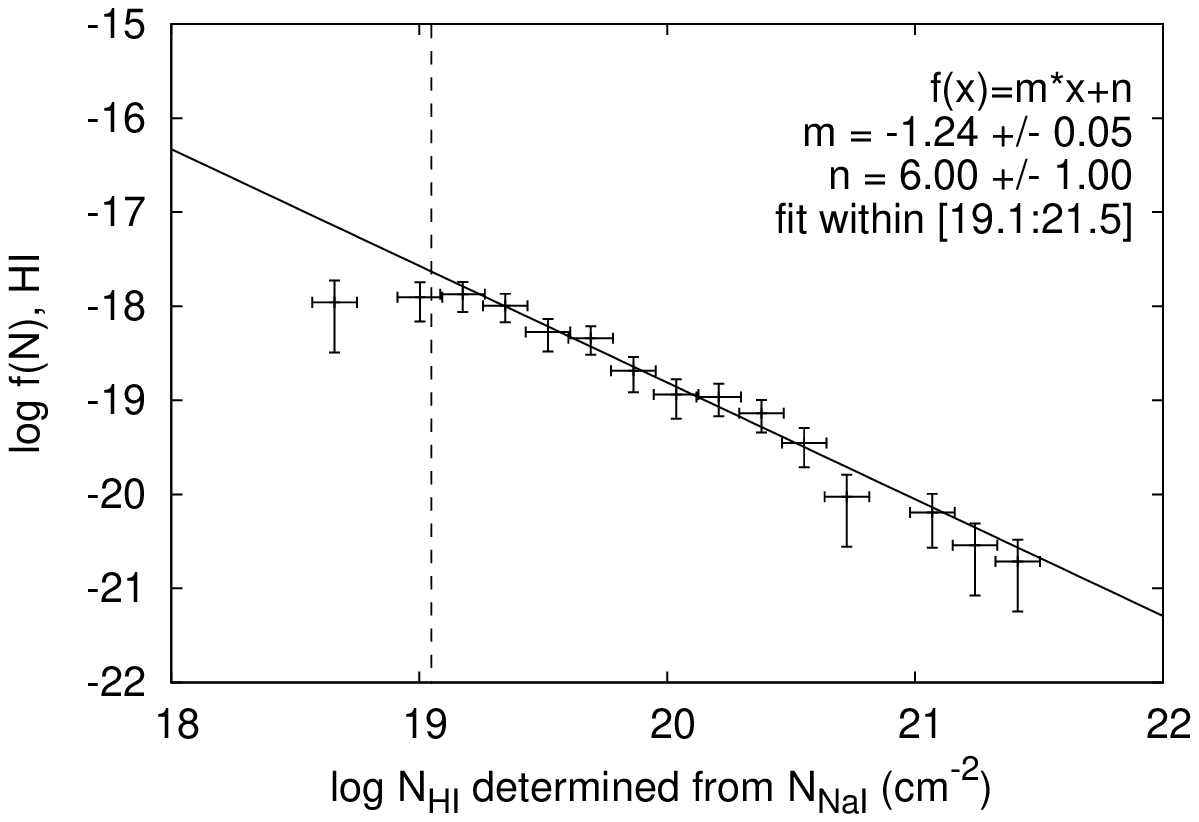}\\[-3ex]
\caption{Converted \ion{H}{i} column density distribution functions.}
\label{figHIcdd}
\end{figure}

We used the conversion formula find by \citet{wakkermathis00} to convert our observed \ion{Ca}{ii} and \ion{Na}{i} column densities into \ion{H}{i} column densities. They used high-resolution absoption and 21-cm emission line data of IVCs and HVCs to find this relations for \ion{Ca}{ii}, \ion{Na}{i}, and \ion{H}{i}. 
These conversions are afflicted with large systematic uncertainties and thus have to be used with caution, because \ion{Ca}{ii} and \ion{Na}{i} with their low ionisation potentials of 12 and 5 eV probably do not represent the dominant ionisation stages in the diffuse ISM. Furthermore, calcium is very easily depleted onto dust grains. Figure\,\ref{figHIcdd} shows the CDD function $f(N)$ of the converted \ion{H}{i} column densities. The solid lines represent the power law fits with $\beta=-1.20 \pm 0.02$ in the range $\log N_\mathrm{CaII}=19 \ldots 25$ (\ion{Ca}{ii}) and $\beta=-1.24 \pm 0.05$ in the range $\log N_\mathrm{NaI}=19.1 \ldots 21.5$ (\ion{Na}{i}).
\citet{petitjeanetal93} have investigated the \ion{H}{i} CDD function of intergalactic QSO absorption line systems with a mean redshift of $z=2.8$ with the help of high spectral resolution data. They find a slope of $\beta=-1.32$ for $\log N_\mathrm{HI}$ between 16 to 22 \citep[see also][]{lehneretal07}. 
It is interesting that the slopes are in general agreement with the statistics of low- and high-redshift QSO absorption line data. It appears that the column density distribution of extraplanar neutral gas structures (i.e., HVCs and their extragalactic analogues) is roughly universal at low and high redshift. 

\begin{figure}
\centering
 \includegraphics[width=0.45\textwidth]{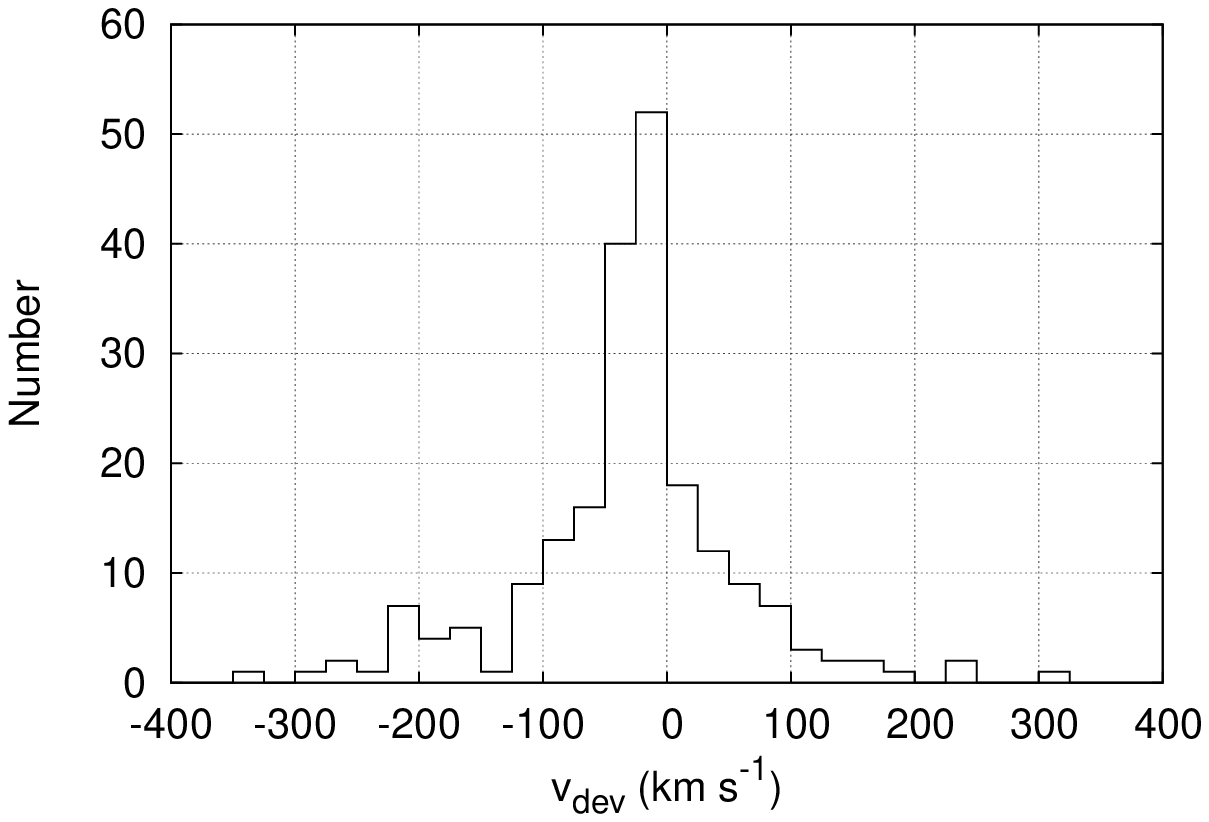}\qquad
 \includegraphics[width=0.45\textwidth]{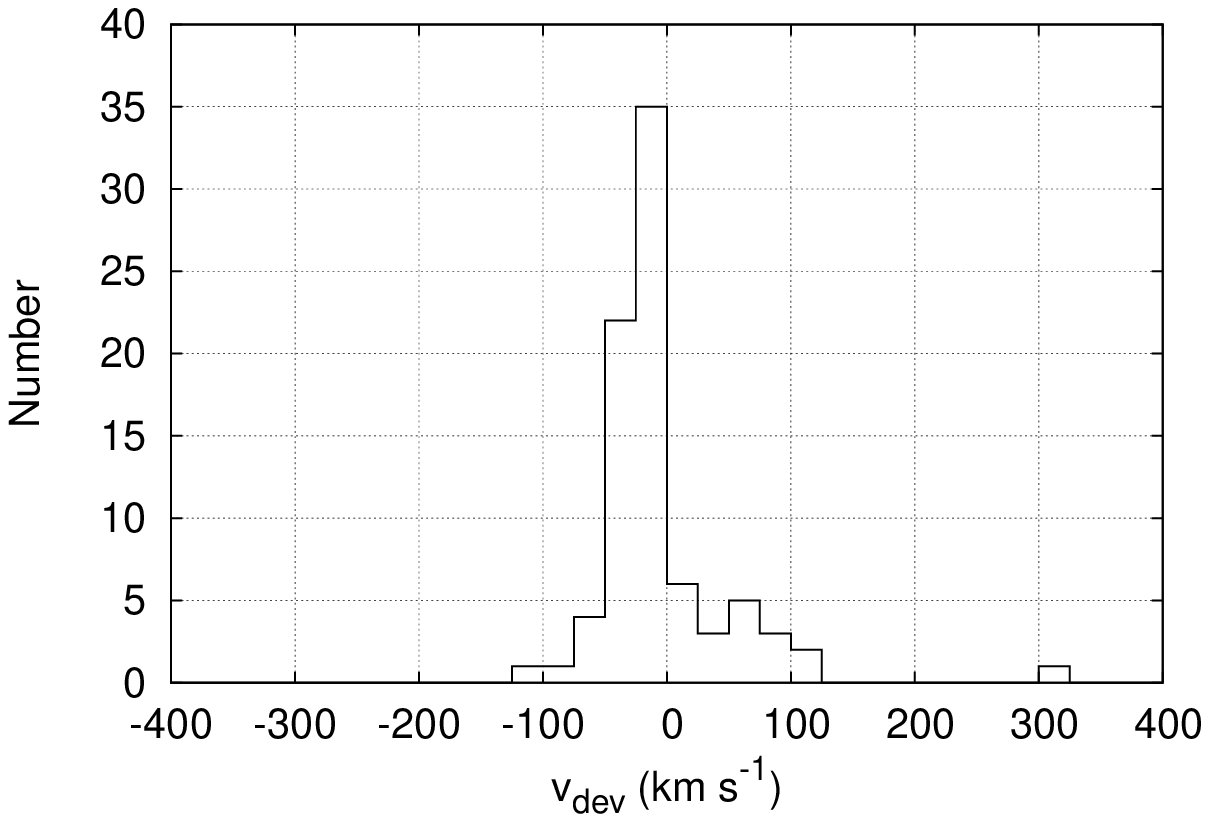}\\[-3ex]
\caption{Distribution of the deviation velocities for the \ion{Ca}{ii} and \ion{Na}{i} absorbers.} 
\label{figvdev}
\end{figure}

Figure\,\ref{figvdev} shows the distribution of the deviation velocities for the \ion{Ca}{ii} and \ion{Na}{i} absorbers in the MW. We observe a slight excess towards negative velocities which probably means that more clouds are infalling to the galactic disk.
For a more detailed statistical analysis see Ben Bekhti et al. (2010, in prep.).

\section{Summary}\label{secsummary}

With our large and sensitive optical absorption and \ion{H}{i} emission line data sample we could show that the MW halo contains a large number of low-column density absorbers producing \ion{Ca}{ii} and \ion{Na}{i} absorption in the spectra of background QSOs. 

In some cases the \ion{Ca}{ii} absorption lines are associated with known intermediate- and high-velocity clouds, but in other cases the observed absorption has no \ion{H}{i} 21-cm counterpart. The observed \ion{Ca}{ii} column density distribution follows a power-law with a slope of $\beta \approx -1.6$. This distribution is similar to the distribution found for intervening \ion{Mg}{ii} systems that trace the gaseous environment of other galaxies at low and high redshift. The results of the converted \ion{H}{i} CDD functions suggest that the halo gas follows the universal power-law of the column density distribution of QSO absorption line systems.  

With radio synthesis telescopes we found small and compact clumps in the direction of all observed sight lines. These observations are in agreement with the observed multiphase structures of clouds in the halos of galaxies. Cold and compact structures are embedded in a warm diffuse envelope.

The presented results show how important measurements at different wavelengths and with different resolutions are to get a complete view of the low-column density halo gas. Analysing several spectral regimes provides us to study the variety of elements, distinct gas phases and structure sizes.

The work which we have done so far represents only the first step in our project. We will go on with the systematically study of the low-column density gaseous environment of the Milky Way and to study the role of the gas for the formation and evolution of our own galaxy. Therefore we are planning to extend our study of extraplanar absorbers using additional high-resolution absorption line data.
We will supplement this sample with \ion{H}{i} single dish and interferometric data. This will allow us to improve our statistical analysis of the distribution and the physical properties of the halo gas (Ben Bekhti et al. 2010, in prep.). The study of absorption from various neutral and weakly ionised species will lead to the determination of metallicities which will help to clarify whether the extraplanar gaseous structures are of Galactic or extragalactic origin.

\acknowledgements The authors would like to thank the Deutsche Forschungsgemeinschaft (DFG) for financial support under the research grants SPP 1177 and KE757/7-1. We are grateful to T. Westmeier for providing the HVC all-sky map.

\bibliography{benbekhti_nadya}

\begin{thebibliography}{}
\expandafter\ifx\csname natexlab\endcsname\relax\def\natexlab#1{#1}\fi
\expandafter\ifx\csname url\endcsname\relax
  \def\url#1{\texttt{#1}}\fi
\expandafter\ifx\csname urlprefix\endcsname\relax\def\urlprefix{URL }\fi
\providecommand{\eprint}[2][]{\url{#2}}

\bibitem[{{Ben Bekhti} et~al.(2008){Ben Bekhti}, {Richter}, {Westmeier}, \&
  {Murphy}}]{benbekhtietal08}
{Ben Bekhti}, N., {Richter}, P., {Westmeier}, T., \& {Murphy}, M.~T. 2008,
  \aap, 487, 583. \eprint{arXiv:0806.3204}

\bibitem[{{Ben Bekhti} et~al.(2009){Ben Bekhti}, {Richter}, {Winkel}, {Kenn},
  \& {Westmeier}}]{benbekhtietal2009}
{Ben Bekhti}, N., {Richter}, P., {Winkel}, B., {Kenn}, F., \& {Westmeier}, T.
  2009, \aap, 503, 483. \eprint{0907.0171}

\bibitem[{{Bouch{\'e}} et~al.(2006){Bouch{\'e}}, {Murphy}, {P{\'e}roux},
  {Csabai}, \& {Wild}}]{boucheetal06}
{Bouch{\'e}}, N., {Murphy}, M.~T., {P{\'e}roux}, C., {Csabai}, I., \& {Wild},
  V. 2006, \mnras, 813. \eprint{astro-ph/0606328}

\bibitem[{{Charlton} et~al.(2000){Charlton}, {Churchill}, \&
  {Rigby}}]{charltonetal00}
{Charlton}, J.~C., {Churchill}, C.~W., \& {Rigby}, J.~R. 2000, \apj, 544, 702.
  \eprint{astro-ph/0002001}

\bibitem[{{Churchill} et~al.(2003){Churchill}, {Vogt}, \&
  {Charlton}}]{churchillvogtcharlton03}
{Churchill}, C.~W., {Vogt}, S.~S., \& {Charlton}, J.~C. 2003, \aj, 125, 98.
  \eprint{arXiv:astro-ph/0210196}

\bibitem[{{Fraternali} et~al.(2007){Fraternali}, {Binney}, {Oosterloo}, \&
  {Sancisi}}]{fraternalietal_07}
{Fraternali}, F., {Binney}, J., {Oosterloo}, T., \& {Sancisi}, R. 2007, New
  Astronomy Review, 51, 95. \eprint{arXiv:astro-ph/0701402}

\bibitem[{{Fraternali} \& {Binney}(2006)}]{fraternaliandbinney06}
{Fraternali}, F., \& {Binney}, J.~J. 2006, \mnras, 366, 449.
  \eprint{arXiv:astro-ph/0511334}

\bibitem[{{Kalberla} et~al.(2005){Kalberla}, {Burton}, {Hartmann}, {Arnal},
  {Bajaja}, {Morras}, \& {P{\"o}ppel}}]{kalberla05}
{Kalberla}, P.~M.~W., {Burton}, W.~B., {Hartmann}, D., {Arnal}, E.~M.,
  {Bajaja}, E., {Morras}, R., \& {P{\"o}ppel}, W.~G.~L. 2005, \aap, 440, 775.
  \eprint{astro-ph/0504140}

\bibitem[{{Kalberla} et~al.(2007){Kalberla}, {Dedes}, {Kerp}, \&
  {Haud}}]{kalberlaetal2007}
{Kalberla}, P.~M.~W., {Dedes}, L., {Kerp}, J., \& {Haud}, U. 2007, \aap, 469,
  511. \eprint{0704.3925}

\bibitem[{{Kalberla} et~al.(2010){Kalberla}, {McClure-Griffiths}, {Pisano},
  {Calabretta}, {Alyson Ford}, {Lockman}, {Staveley-Smith}, {Kerp}, {Winkel},
  {Murphy}, \& {Newton-McGee}}]{kalberlaetal_2010}
{Kalberla}, P.~M.~W., {McClure-Griffiths}, N.~M., {Pisano}, D.~J.,
  {Calabretta}, M.~R., {Alyson Ford}, H., {Lockman}, F.~J., {Staveley-Smith},
  L., {Kerp}, J., {Winkel}, B., {Murphy}, T., \& {Newton-McGee}, K. 2010, ArXiv
  e-prints. \eprint{1007.0686}

\bibitem[{{Lehner} et~al.(2007){Lehner}, {Savage}, {Richter}, {Sembach},
  {Tripp}, \& {Wakker}}]{lehneretal07}
{Lehner}, N., {Savage}, B.~D., {Richter}, P., {Sembach}, K.~R., {Tripp}, T.~M.,
  \& {Wakker}, B.~P. 2007, \apj, 658, 680. \eprint{arXiv:astro-ph/0612275}

\bibitem[{{McClure-Griffiths} et~al.(2009){McClure-Griffiths}, {Pisano},
  {Calabretta}, {Ford}, {Lockman}, {Staveley-Smith}, {Kalberla}, {Bailin},
  {Dedes}, {Janowiecki}, {Gibson}, {Murphy}, {Nakanishi}, \&
  {Newton-McGee}}]{mcClureetal2009}
{McClure-Griffiths}, N.~M., {Pisano}, D.~J., {Calabretta}, M.~R., {Ford},
  H.~A., {Lockman}, F.~J., {Staveley-Smith}, L., {Kalberla}, P.~M.~W.,
  {Bailin}, J., {Dedes}, L., {Janowiecki}, S., {Gibson}, B.~K., {Murphy}, T.,
  {Nakanishi}, H., \& {Newton-McGee}, K. 2009, \apjs, 181, 398.
  \eprint{0901.1159}

\bibitem[{{Petitjean} et~al.(1993){Petitjean}, {Webb}, {Rauch}, {Carswell}, \&
  {Lanzetta}}]{petitjeanetal93}
{Petitjean}, P., {Webb}, J.~K., {Rauch}, M., {Carswell}, R.~F., \& {Lanzetta},
  K. 1993, \mnras, 262, 499

\bibitem[{{Richter} et~al.(2001){Richter}, {Sembach}, {Wakker}, {Savage},
  {Tripp}, {Murphy}, {Kalberla}, \&
  {Jenkins}}]{richter_sembach_wakker_savage_tripp_kalberlaetal01}
{Richter}, P., {Sembach}, K.~R., {Wakker}, B.~P., {Savage}, B.~D., {Tripp},
  T.~M., {Murphy}, E.~M., {Kalberla}, P.~M.~W., \& {Jenkins}, E.~B. 2001, \apj,
  559, 318. \eprint{arXiv:astro-ph/0105466}

\bibitem[{{Richter} et~al.(2005){Richter}, {Westmeier}, \&
  {Br{\"u}ns}}]{richterwestmeierbruens05}
{Richter}, P., {Westmeier}, T., \& {Br{\"u}ns}, C. 2005, \aap, 442, L49.
  \eprint{astro-ph/0509585}

\bibitem[{{Sancisi} et~al.(2008){Sancisi}, {Fraternali}, {Oosterloo}, \& {van
  der Hulst}}]{sancisietal2008}
{Sancisi}, R., {Fraternali}, F., {Oosterloo}, T., \& {van der Hulst}, T. 2008,
  \aapr, 15, 189. \eprint{0803.0109}

\bibitem[{{Sembach} et~al.(2003){Sembach}, {Wakker}, {Savage}, {Richter},
  {Meade}, {Shull}, {Jenkins}, {Sonneborn}, \&
  {Moos}}]{sembach_wakker_savage_richter_etal03}
{Sembach}, K.~R., {Wakker}, B.~P., {Savage}, B.~D., {Richter}, P., {Meade}, M.,
  {Shull}, J.~M., {Jenkins}, E.~B., {Sonneborn}, G., \& {Moos}, H.~W. 2003,
  \apjs, 146, 165. \eprint{arXiv:astro-ph/0207562}

\bibitem[{{Wakker} \& {Mathis}(2000)}]{wakkermathis00}
{Wakker}, B.~P., \& {Mathis}, J.~S. 2000, \apjl, 544, L107.
  \eprint{astro-ph/0010045}

\bibitem[{{Winkel} et~al.(2010){Winkel}, {Kalberla}, {Kerp}, \&
  {Fl{\"o}er}}]{winkeletal2010}
{Winkel}, B., {Kalberla}, P.~M.~W., {Kerp}, J., \& {Fl{\"o}er}, L. 2010, \apjs,
  188, 488. \eprint{1005.4604}

\end{thebibliography}

\end{document}